\documentclass[twocolumn,prb,preprintnumbers,showpacs,amsmath,amssymb]{revtex4}
\usepackage{graphicx}
\usepackage{dcolumn}
\usepackage{txfonts}
\usepackage{xcolor}
\usepackage{amssymb}
\usepackage{amsmath}
\usepackage{latexsym}
\usepackage{epsfig}
\usepackage{amsbsy}
\usepackage{array}
\usepackage{tabularx}

\begin{document}
\title [] {Effects of dilute substitutional solutes on carbon in $\alpha$-Fe:
interactions and associated carbon diffusion from first-principles calculations}

\author{Peitao Liu}
\author{Weiwei Xing}
\author{Xiyue Cheng}
\author{Dianzhong Li}
\author{Yiyi Li}
\author{Xing-Qiu Chen}
\email[Corresponding author: ]{xingqiu.chen@imr.ac.cn}

\affiliation{Shenyang National Laboratory for Materials
Science, Institute of Metal Research, Chinese Academy of Sciences,
Shenyang 110016, China}

\date{\today}

\begin{abstract}
By means of first-principles calculations coupled with the kinetic
Monte Carlo simulations, we have systematically investigated the
effects of dilute substitutional solutes on the behaviors of carbon
in $\alpha$-Fe. Our results uncover that: ($i$) Without the Fe
vacancy the interactions between most solutes and carbon are
repulsive due to the strain relief, whereas Mn has a weak attractive
interaction with its nearest-neighbor carbon due to the local
ferromagnetic coupling effect. ($ii$) The presence of the Fe vacancy
results in attractive interactions of all the solutes with carbon.
In particular, the Mn-vacancy pair shows an exceptionally large
binding energy of -0.81 eV with carbon. ($iii$) The alloying
addition significantly impacts the atomic-scale concentration
distributions and chemical potential of carbon in the Fe matrix.
Among them, Mn and Cr increase the carbon chemical potential whereas
Al and Si reduce it. ($iv$) Within the dilute scale of the alloying
solution, the solute concentration and temperature dependent carbon
diffusivities demonstrate that Mn has a little impact on the carbon
diffusion whereas Cr (Al or Si) remarkably retards the carbon
diffusion. Our results provide certain implication for better
understanding the experimental observations related with the carbon
solubility limit, carbon micro-segregation and carbide
precipitations in the ferritic steels.

\end{abstract}


\maketitle

\section{Introduction}

Carbon is one of the most common interstitial atoms in Fe-based
alloys. Its addition can significantly improve the strength and
hardness of steels by the solution strengthening and carbides
precipitation strengthening.\cite{Abbaschian R} Carbon is also
incorporated into the surface layers by carburizing to enhance the
mechanical properties of steels. Besides carbon, many other alloying
elements (i.e., 3$d$, 4$d$ and 5$d$ transition metals) are also
added to improve the performances of the steels. Undoubtedly, the
alloying elements would inevitably interact with carbon. Their
additions would not only result in the lattice distortion and the
local strain field due to the size factor of solutes, but also
induce chemical or electronic effects on the soluble carbon atoms.
They would trap or repel carbon, thereby affecting the behaviors of
carbon in steels, such as the carbon solubility limit
\cite{Saitoh-1,Saitoh-2}, micro-segregation \cite{Suzuki,Lu YP},
diffusion \cite{diffusion-1,diffusion-2,diffusion-3,diffusion-4} and
carbides precipitations \cite{carbides-1,carbides-2,carbides-3}.

In comparison with other alloying elements in steels, carbon is
lighter in mass, smaller in size and less in the electronic valence
number. Therefore, it is highly difficult to experimentally identify
the carbon's behaviors in steels within the atomic scale. In
particular, since in $\alpha$-Fe the carbon solubility is very
limited and its kinetics are very slow, it would require a long time
to yield a true equilibrium. Although many studies have been
performed to analyze the solute-carbon interactions, most
understandings were derived from the sophisticated mechanical
spectroscopic measurements. From the viewpoint of atomistic
interactions, many questions remain open. For instance, through the
internal friction measurements combined with the infrared analysis
of carbon, Saitoh \emph{et al.} \cite{Saitoh-2} reported that in
$\alpha$-Fe Mn and Cr hardly altered the carbon solubility limit,
whereas P and Si enhance it. The reasons of these behaviors still
remain unclear. It is also well known that the carbon
micro-segregation is more serious in high Mn steels, as observed by
both Suzuki \emph{et al.} \cite{Suzuki} and Lu \emph{et al.}\cite{Lu
YP}. However, the in-depth mechanism has not been resolved.
Therefore, it would be highly desirable to elucidate the atomistic
interactions between solutes and carbon in $\alpha$-Fe, which could
provide better understanding for the phase equilibria, phase
diagrams and mechanical as well as physical properties
\cite{Sozinov}.

To date, \emph{ab initio} calculation based on the density
functional theory (DFT) has been demonstrated to be a powerful tool
to accurately evaluate the atomic interactions and understand the
basic atomic phenomena involved. For instance, Jiang and Carter
\cite{Jiang-1, Jiang-2} investigated the carbon dissolution and
diffusion in the ferrite and austenite, as well as the carbon
adsorption and diffusion into Fe (110) and Fe (100) surfaces from
first-principles calculations. Domain \emph{et al.} \cite{Domain}
discussed the interactions of one C atom with an Fe vacancy, another
C atom, and the self-interstitial atoms in $\alpha$-Fe. They
concluded that an Fe vacancy could bind two carbon atoms at most and
the carbon-carbon interactions were revealed to be mostly repulsive.
Afterwards, the interactions of an Fe vacancy with multiple C atoms
in $\alpha$-Fe were investigated in details by Ohnuma \emph{et al.}
\cite{Ohnuma}, who claimed that an Fe vacancy would bind four carbon
atoms at most, but the system with a vacancy binding three C atoms
was the most energetically stable. Yan \emph{et al.} \cite{Yan} even
studied the interactions of C-N and C-N-vacancy in $\alpha$-Fe. They
found that both C and N atoms would separate away from each other as
far as possible in steels. Furthermore, utilizing the derived
\emph{ab initio} binding energies, Becquart \emph{et al.}
\cite{Becquart} constructed the so-called Fe-C potential within the
framework of Embedded Atom Method (EAM), determining the interaction
of the carbon atoms with a screw dislocation. By combining
first-principles calculations with the kinetic Monte Carlo (kMC)
method, the Si impacts on the carbon diffusivity in $\alpha$-Fe have
been investigated by Simonovic \emph{et al.}\cite{Simonovic}. Very
recently, Bakaev \emph{et al.}\cite{Bakaev} explored the
interactions of some minor alloying elements in ferritic steels with
interstitial carbon atom using \emph{ab initio} calculations. They
found that Mn exhibits peculiar behavior. In difference from other
elements, Mn shows attractive interactions with carbon in the first-
and second-nearest-neighbor sites\cite{Bakaev}.

Although some interactions ({\em i.e.}, vacancy-carbon (or
transition metal solutes), carbon-carbon (or nitrogen), and
solute-solute ones) have already been investigated theoretically,
the systematic studies on the interactions between substitutional
solutes and carbon in $\alpha$-Fe are rare, not to mention a unified
understanding. The work we present here is intended to contribute to
such an understanding from the perspective of state-of-the-art
\emph{ab initio} calculations by systematically elucidating the
solute-C interactions, solute-vacancy-C interactions, and the
impacts of the substitutional solutes on the carbon's distribution,
chemical potential and diffusion. These results would be definitely
useful for the in-depth understanding of the carbon solubility
limit, carbides precipitations and the occurrence of carbon
micro-segregation.

\section{Methodology}
\subsection{First-principles calculations and binding energies}

Our calculations were based on the framework of density functional
theory (DFT) \cite{Hohenberg, Kohn} as implemented in the Vienna
\emph{Ab initio} Simulation Package (VASP)\cite{Kresse-1, Kresse-2}.
All calculations were performed using the projector augmented wave
(PAW) \cite{PAW} potentials and the generalized-gradient
approximation (GGA) within the Perdew-Burke-Ernzerhof (PBE)
\cite{PBE} exchange-correlation function, which has been proved to
provide an accurate description of the magnetic and energetic
properties of Fe bulk phases\cite{Singh}. We used a Fermi smearing
of electronic occupancy with a width of 0.05 eV, and a plane wave
cut-off energy of 500 eV, which has been found to be sufficient for
precise energetics for all the elements considered in present work.
Spin polarized calculations were performed by considering the
ferromagnetic ordering of Fe. The ion relaxations were performed at
constant volume rather than at constant pressure, since the former
one was found more suitable for the bcc type cells in the impurities
calculations\cite{Klymko}. The quasi-Newton algorithm was used to
capture the minimum energy. The local magnetic moments and local
density of states were calculated through the evaluation of the spin
density within the Wigner-Seitz spheres around nuclei, the values of
which adopted here were the recommended ones for the VASP code.

All the binding energy calculations were based on a
3$\times$3$\times$3 bcc unit cell, which contains 54 Fe atoms in the
defect-free state. A 5$\times$5$\times$5 $\vec{k}$-mesh grid
generated by the Monkhorst-Pack scheme was used to sample the
Brillouin zone, which has been revealed to be large enough to
calculate the formation and binding energies for carbon with point
defects \cite{Jiang-1,Jiang-2}. Certainly, we have doubly checked
the supercell convergence by computing the carbon solution enthalpy
and the Fe vacancy formation energy using a larger
4$\times$4$\times$4 supercell with 3$\times$3$\times$3
$\vec{k}$-mesh grid. It is found that the energy differences between
these two supercells are both less than 0.01 eV. To aid the
computational efficiency, the projections operators were evaluated
in real space because of the larger number of atoms in a supercell
\cite{Counts}. In agreement with the published results
\cite{Jiang-1,Jiang-2}, carbon was found to occupy the most stable
octahedral interstitial site (\emph{o}-site). Our calculations also
revealed that the transition-metal elements, Al and Si would
substitute Fe site due to their comparatively large atomic size. For
the $\alpha$-Fe, the calculated lattice parameter and magnetic
movement are 2.83 \AA\ and 2.2 $\mu_\text{B}$/atom, respectively, in
good agreement with the experimental data of 2.86 \AA\ and 2.2
$\mu_\text{B}$/atom \cite{Kittel C}.

The binding energies are used to evaluate the interactions. In cases
where the defect cluster contains two point defects, the binding
energies are defined as follows:
\begin{equation}
E_{b}^{i,j}=\frac{1}{m}[{E}_{D-(i+j)}-{E}_{D-i}-{E}_{D-j}+{E}_{ref}]
\label{eq:ee1}
\end{equation}
where $E_{D-i}$ and  $E_{D-j}$ are the total energy of the supercell
with the point defect \emph{i} and \emph{j}, respectively,
$E_{D-(i+j)}$ the energy of the supercell containing both point
defect \emph{i} and \emph{j}, and $E_{ref}$ the energy of the
defect-free supercell, which is used to balance the number of the Fe
atoms. \emph{m} is the multiplicity considering the finite size and
the periodicity of the cell \cite{Simonovic}. For instance, if the
substitutional atom is placed at [000] site and carbon is at [3/2
00] site, the substitutional atom would interact with carbon twice,
and thus $m$ equals to 2.

If the defect cluster contains three point defects, two different
binding energies: the total binding energy and the incremental
binding energy would be defined.\cite{Counts} The total binding
energy \cite{Counts}, representing the stability of the system with
respect to the isolated defect, is defined as the energy difference
between the supercell with the triple defects and the three
supercells with individual point defect,
\begin{equation}
E_{b}^{i,j,k}={E}_{Triple-(i+j+k)}-{E}_{D-i}-{E}_{D-j}-{E}_{D-k}+2{E}_{ref}
\label{eq:ee2}
\end{equation}
where ${E}_{Triple-(i+j+k)}$ is the energy of the supercell
containing all three point defects. The incremental binding energy
\cite{Counts} is defined as the energy difference between the
supercell with the triple defects and the supercells with a single
point defect and a defect pair,
\begin{equation}
E_{b}^{i,j+k}={E}_{Triple-(i+j+k)}-{E}_{D-i}-{E}_{Pair-(j+k)}+{E}_{ref}
\label{eq:ee3}
\end{equation}
where ${E}_{Pair-(j+k)}$ is the energy of the supercell containing a
pair of defects. It should be noted that since the considered
interaction distance within the defect cluster is short, the
multiplicity $m$ mentioned in Eq. (\ref{eq:ee1}) is 1 in Eqs.
(\ref{eq:ee2})-(\ref{eq:ee3}). According to the definition in Eqs.
(\ref{eq:ee1})-(\ref{eq:ee3}), a negative binding energy indicates a
favorable and attractive interaction between defects, while a
positive binding energy refers to an unfavorable and repulsive
interaction. This convention will be used to explore and explain all
the interactions in various configurations discussed below.

\subsection{Computations of the influences of dilute
solutes on the carbon's distribution and chemical
potential}

At the dilute concentration, the substitutional solutes would
randomly distribute in the $\alpha$-Fe matrix where the individual
solute atoms are far apart from each other. The carbon atoms, which
diffuse much faster than the substitutional solutes, will arrange
themselves around the solutes \cite{Simonovic}. In the limit of low
solute and carbon concentration and under the condition of
thermodynamical equilibrium, the carbon-carbon (C-C) and
solute-solute ($M$-$M$) interactions can thus be neglected. In terms
of the model proposed by Simonovic \cite{Simonovic}, the probability
${f}_{Rs}$ that a carbon atom is present at a certain distance $Rs$
from solute $M$ can be expressed as follows,
\begin{equation}
f_{Rs}={f}_{\infty}\text{exp} \Big(\frac{-E_b^{M,C}(Rs)}{{k}_{B}{T}}\Big)
\label{eq:ee5}
\end{equation}
where $E_b^{M,C}(Rs)$ is the solute-C binding energy at the distance
of $Rs$. ${k}_{B}$ and $T$ represent the Boltzmann constant and the
absolute temperature, respectively. As the distance approaches
$Rs_{max}$, where the solute-C interactions vanish, the fraction of
octahedral sites filled with carbon atoms can be assumed to be a
constant value of ${f}_{\infty}$. The carbon concentration
${C}_{\text{C}}$ with respect to the bcc lattice is then given as
follows \cite{Simonovic},
\begin{equation}
{C}_{\text{C}}={C}_M \sum_{s=1}^{max} {n}_{s} {f}_{Rs} +
\Big(3-{C}_M \sum_{s=1}^{max} {n}_{s}\Big) {f}_{\infty}
\label{eq:ee6}
\end{equation}
where the first sum concerns all the carbon atoms within the
interaction range $s_{max}$ of the solute atom, and the value ``3''
in the second term denotes three octahedral positions per atom in
the bcc lattice. ${C}_M$ is the concentration of the solute $M$ on
the bcc lattice sites, and $\emph{n}_{s}$ is the number of the
octahedral sites on the shell $s$ of the solute. Practically, the
carbon concentration ${C}_{\text{C}}$ is fixed. Combining Eq.
(\ref{eq:ee5}) and Eq. (\ref{eq:ee6}) the carbon fraction beyond the
interaction range \cite{Simonovic} can be derived as follows,
\begin{equation}
f_{\infty}=\frac{{C}_{\text{C}}}
{{C}_{M}\sum_{s=1}^{max}{n}_{s} \text{exp}(\frac{-E_b^{M,C}(Rs)}{{k}_{B}T})+
(3-{C}_{M}\sum_{s=1}^{max}{n}_{s})}
\label{eq:ee7}
\end{equation}
According to the theory of the ideal solution, the chemical
potential of carbon far away from the solute atom can be
approximated as \cite{Simonovic},
\begin{equation}
\begin{split}
{\mu}_{\text{C}}&=\frac{dF}{{dC}_{\text{C}}}\\
&=\frac{dF}{{df}_{\infty}}\frac{{df}_{\infty}}{{dC}_{\text{C}}}\\
&\approx\mu_{\text{C}}^{0}+3{k}_B{T} \text{ln} \Big(\frac{f_{\infty}}{1-f_{\infty}}\Big)\frac{1}{3}\\
&=\mu_{\text{C}}^{0}+{k}_B{T} \text{ln}({f}_{\infty})
\label{eq:ee8}
\end{split}
\end{equation}
where $F$ is the free energy and $\mu_{\text{C}}^{0}$ is the
reference chemical potential. It can be inferred from the Eq.
(\ref{eq:ee8}) that the chemical potential of carbon
${\mu}_\text{C}$ will reduce when ${f}_{\infty}$ decreases.

 \subsection{Computations of kMC simulations}

To evaluate the migration energy barrier of carbon in $\alpha$-Fe in
the presence of vacancy, solute $M$ or solute-vacancy pair, the
climbing-image nudged elastic band (CI-NEB)\cite{NEB,NEB-2} method
was employed. This method provides a way to find an minimum energy
pathway (MEP) given the initial and final states of a process.
During the NEB calculations, the images were kept separated using a
spring force constant of 5 eV/\AA\, and then relaxed using a
conjugate gradient algorithm until the maximum force acting on each
atom was less than 0.01 eV/\AA. Note that in the cases where the
carbon migrates in the presence of the solutes, we have computed the
migration energy barriers of the carbon for all the possible
diffusion pathways (see Table. \ref{tab:Table3}), as the direction
in which carbon will choose to jump depends significantly on its
surrounding environment. These barriers can be further used as
inputs for the following kMC simulations.

Carbon diffusivity in the presence of the dilute solutes has been
further derived by the kMC method, which can be used to simulate the
dynamic properties within a larger time-scale because the time step
is updated during the simulations.\cite{Kalos} In the kMC
simulations, we have employed a very large simulated box (a
30$\times$30$\times$30 bcc unit cell with the periodical boundary
condition). The bcc lattice positions were all occupied either by Fe
or solute $M$ atoms. The solutes were randomly distributed according
to their atomic concentration. The solute-solute distance was kept
far beyond their interactions to form an approximate dilute
environment. It needs to be emphasized that the diffusions of
solutes and Fe are neglected here as their diffusions are extremely
lower than that of carbon in $\alpha$-Fe \cite{Simonovic}. For
carbon atom in pure $\alpha$-Fe, the diffusion prefactor $D_0$ and
the corresponding diffusion migration energy $\Delta E$ were
measured to be $D_0 =6.61\times10^{-7}$ m$^2$/s and $\Delta E =
0.83$ eV, respectively\cite{Pascheto}. But the experimental
self-diffusion data of Fe were $D_0 =6.8\sim27.7\times10^{-4}$
m$^2$/s and $\Delta E = 2.95\sim3.10$ eV \cite{Mehrer}. Even at 1000
K, the carbon diffusivity is three orders of magnitude greater than
the self-diffusivity of Fe. This fact is also similar for most of
other substitutional solutes \cite{Huang}. Within this context, it
is safe to assume that the dilute substitutional solutes and Fe
atoms do not diffuse as compared to the carbon atoms.

Technically, only one single carbon atom is considered in our kMC
simulations. It is randomly located at the octahedral site at the
initial status, and then is allowed to jump to the neighboring
\emph{o}-sites according to the probability rates which are computed
within the framework of the transition state theory (TST)
\cite{TST}. The energy barrier of each jump is calculated through
CI-NEB method \cite{NEB,NEB-2}. The transition time between two
consecutive jumps is determined by the probability rates. The
detailed steps of our kMC simulations are further compiled as
follows\cite{Kalos},

($i$)  For each jump of the carbon atom, the probability rate
$\gamma_i$ is calculated by ${\gamma}_i=\nu_0 \text{exp}
\Big(\frac{-{E}_{b}^{i}}{{k}_{B}{T}}\Big)\quad (i=1,2,3,4)$, where
$E_{b}^{i}$ is the energy barrier for the jump direction $i$ and
$\nu_0$ is the jump attempt frequency, which was calculated to be
$\nu_0 = 6.476\times10^{13}  s^{-1}$ based on the Einstein
approximation \cite{Simonovic}. Note that the maximum number of $i$
is 4 for each jump because only four jump directions can be chosen
from the current \emph{o}-site to the nearest neighboring
\emph{o}-site in the bcc lattice.

($ii$) The total probability rate $F= \sum_{i=1}^4\gamma_{i}$ and
the relative probability of each event $F_i =
\sum_{j=1}^i\gamma_{j}/F$ are computed.

($iii$) The jump event $i$ is selected by obeying
$\sum_{k=1}^{i-1}F_{k}< \mu \leq \sum_{k=1}^i F_k$, where $\mu$ is a
uniform random number $\mu \in$[0, 1].

($iv$)  Meanwhile, the time elapsed for the current time step is
calculated by ${\triangle}t = - \text{ln}(\xi)/ F$ based on the
residence-time algorithm\cite{residence-time algorithm}, where $\xi$
is another random number distributed uniformly in [0,1]. Then the
physical time increases $t = t+{\triangle}t$.

The steps ($i$) to ($iv$) are repeated until the physical time
reaches the specified time (about 20,000 jump steps) after the
carbon atom has moved a certain distance $R$ away from the original
position. Finally, the carbon diffusivity $D$ can be derived from
the Einstein relation \cite{Einstein},
\begin{equation}
<R^2>=6Dt
\label{eq:ee10}
\end{equation}
where $<R^2>$ denotes the mean square displacement of carbon
obtained by averaging over long time and repeated simulations. The
prefactor $D_0$ and the migration energy $\Delta E$ for the carbon
diffusion can then be extracted by the empirical Arrhenius form of
the diffusion equation \cite{Einstein},
\begin{equation}
D=D_0 \text{exp} \Big(\frac{-{\Delta}E}{{k}_B{T}}\Big)
\label{eq:ee11}
\end{equation}

\section{Results and discussions}
\subsection{Solute-carbon ($M$-C) interactions}

\begin{table*}
\footnotesize \caption {Local magnetic moments
($\mu_{\text{B}}$/atom) for the $1nn$ configuration of
Fe$_{53}M_1$C$_1$ and for the $cfg^5$ of Fe$_{52}M_1$C$_1$ ($M$=3$d$
transitional metal elements). The $inn$ represents the $i$th
nearest-neighbor site of Fe relative to C.}
\begin{ruledtabular}
\begin{tabular}{cccccccccccccc}
$M$-C&    $M$  &   C   &   Fe$1nn$ &   Fe$2nn$ &   Fe$3nn$ & & & $M$-Vac-C&  $M$   &   C   &   Fe$1nn$ &   Fe$2nn$ &   Fe$3nn$ \\
\hline
Ti      &    -0.52    &   -0.09    &   1.46    &   2.10    &   2.38    & & &    Ti      & -0.74  & -0.13      &   2.15    &   2.51    &   2.52    \\
V       &   -0.71     &   -0.08    &   1.53    &   2.14    &   2.39   & &   &   V       & -1.34  &   -0.15   &   1.71    &   2.03    &   2.04    \\
Cr      &   -0.66     &   -0.07    &   1.56    &   2.15    &   2.39    & & &   Cr     &   -2.11 & -0.12     &   0.93    &   2.09    &   2.11    \\
Mn    &   0.68       &    -0.11   &   1.68    &   2.21    &   2.39    & & &   Mn    &  -2.67  &   -0.15   &   1.72    &   2.02    &   2.03    \\
Fe      &   1.67      &  -0.14     &   1.67    &   2.20    &   2.38    & & &   Fe      &   2.42   &   -0.15   &   1.70    &   2.03    &   2.02    \\
Co    &   1.22       &    -0.13    &   1.67    &   2.21    &   2.38   & & &   Co     &   1.70    &   -0.15   &   1.73    &   2.03    &   2.03    \\
Ni     &   0.57       &    -0.12    &   1.60    &   2.17    &   2.36    & & &  Ni     &   0.92    &   -0.15   &   1.73    &   2.02    &   2.03    \\
Cu    &   0.08       &    -0.12    &   1.47    &   2.10    &   2.35    & & &  Cu     &   0.18    &   -0.15   &   1.73    &   2.03    &   2.03    \\
\end{tabular}
\end{ruledtabular}
\label{tab:Table1}
\end{table*}

\begin{figure}[b!]
\begin{center}
\includegraphics[width=0.35\textwidth]{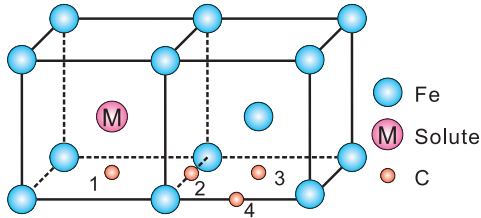}
\end{center}
\caption{Solute \emph{M}-C configurations. The C atom's site labeled
\emph{i} represents the first- to the fourth-nearest-neighbor (1$nn$
to 4$nn$) octahedral site relative to the solute $M$ in the bcc
lattice.  } \label{fig:fig1}
\end{figure}

\begin{figure}[b!]
\begin{center}
\includegraphics[width=0.45\textwidth]{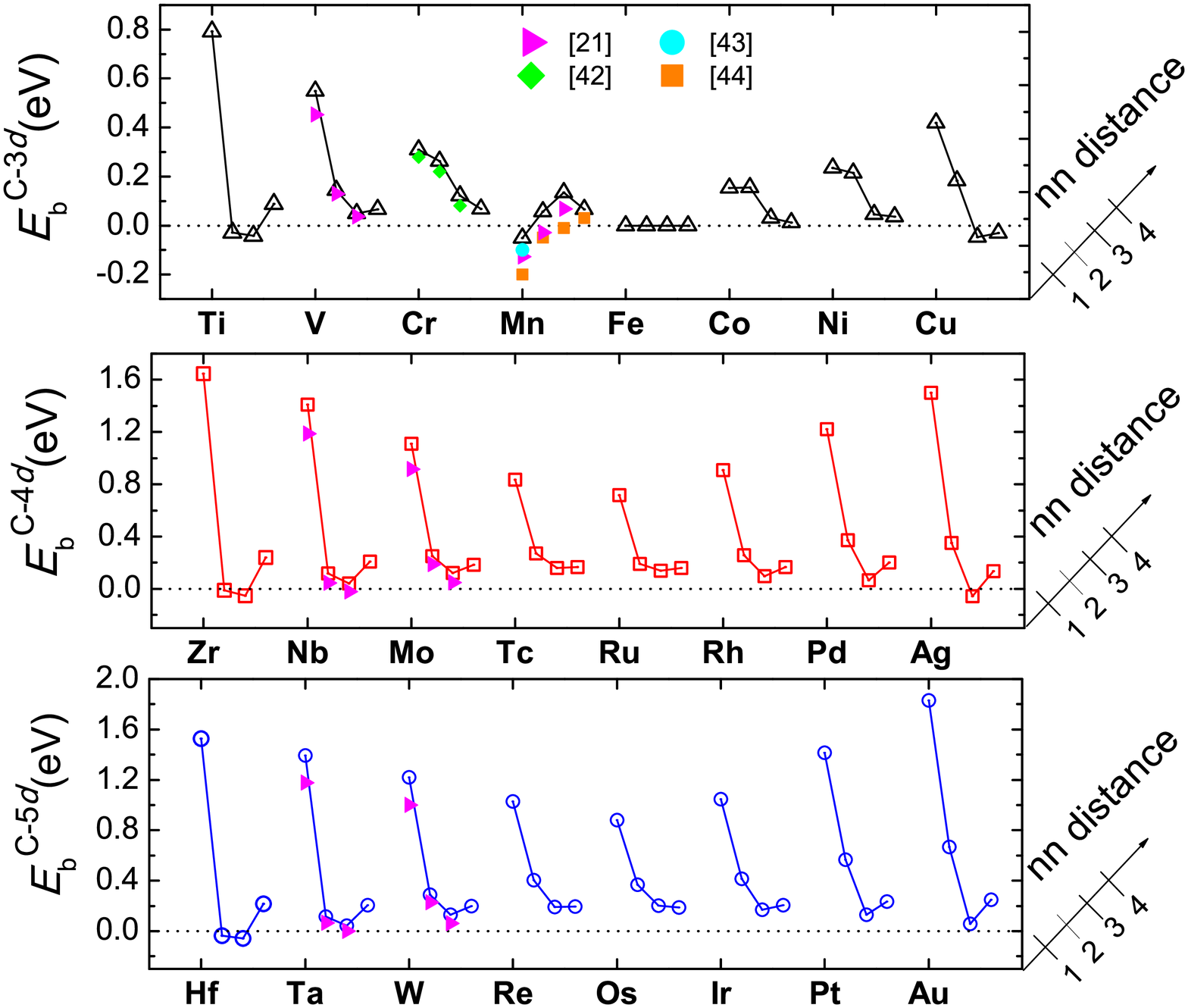}
\end{center}
\caption{Solute $M$-C binding energies for the 3$d$ (top panel),
4$d$ (middle panel), and 5$d$ (bottom panel) elements with C in the
1$nn$ to 4$nn$ octahedral site relative to the solute $M$. }
\label{fig:fig2}
\end{figure}

In order to calculate the solute-C binding energies, we have
constructed the model (Fig. \ref{fig:fig1}), where the solute $M$
substitutes an Fe atom and the carbon is inserted in different
neighboring \emph{o}-sites with respect to the solute \emph{M}.
Figure \ref{fig:fig2} shows the binding energies between all the
transition-metal elements from the groups 4 to 11 on the periodic
table and the carbon at the $1nn$ (first-nearest-neighbor) to the
$4nn$ (fourth-nearest-neighbor) site. It can be seen that the
solute-C interactions for all the 4$d$ and 5$d$ rows elements almost
exhibit a similar tendency: the 1$nn$ site is the least stable one
for carbon to stay due to the largest repulsive interactions. As the
distance increases, the repulsive interactions decrease sharply.
Carbon seems to prefer to stay at the 3$nn$ site since this position
is energetically lowest. However, their solute-C (at 3$nn$) binding
energies are still positive except for Zr, Hf and Ag which have a
very weak attractive interaction with carbon. With increasing the
distance to the 4$nn$ site, the binding energies become positive,
again.

\begin{figure}[b!]
\begin{center}
\includegraphics[width=0.40\textwidth]{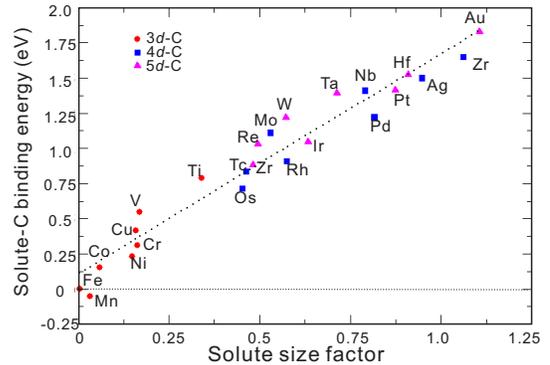}
\end{center}
\caption{Solute $M$-C binding energies for the 1$nn$ configuration
as a function of the solute size factor (The size factor of solute
$M$ in the Fe matrix is defined as, $\Omega
_{sf}^M=\frac{\Omega_M-\Omega _{Fe}}{\Omega _{Fe}}$, where $\Omega
_M$ and $\Omega_{Fe}$ are the volume of supercell with the solute
$M$ and defect-free supercell, respectively. See more details in the
Ref. \cite{Olsson}.). } \label{fig:fig3}
\end{figure}

The repulsion between all the 4\emph{d} and 5\emph{d}
transition-metal elements and carbon is no surprise. Since these
elements are larger in size than Fe, their insertions would
definitely result in a large local strain. The weak attraction for
Zr, Hf and Ag binding with C at the 3$nn$ site can also be explained
by the elastic effects because the 3$nn$ octahedral interstices
impacted by these elements are less flattened than those in the pure
$\alpha$-Fe crystal structure. Interestingly, it has been found that
those 4\emph{d} or 5\emph{d} solute-C binding energies on the 1$nn$
shell can be nearly linearly correlated with the size factor (as
defined in Ref. \cite{Olsson}) of those solutes, as illustrated in
Fig. \ref{fig:fig3}. With the increasing solute size factor, the
repulsive interaction becomes more obvious. Hence, it can be
inferred that the individual 4\emph{d} and 5\emph{d} rows element
interacts with carbon atom mainly through the strain relief.

However, the binding energies for the 3$d$ row elements exhibit a
much more complicated behavior (see Fig. \ref{fig:fig2}), because
their \emph{M}-C interactions depend not only on the solute size
factors but also on the stronger magnetic couplings around Fe. Ti
and V experience the antiferromagnetic coupling with their 1$nn$ Fe
atoms, whereas Cu has a weak ferromagnetic coupling with the 1$nn$
Fe atoms (Table. \ref{tab:Table1}). However, it needs to be
emphasized that, although the magnetic coupling exists for them,
their solute-C interactions are mainly dominated by the strain
relief because of their relatively large solute size factors (Fig.
\ref{fig:fig3}). Therefore, these elements show the similar behavior
with the the 4\emph{d} or 5\emph{d} elements of comparable sizes.
In contrast, for the intermediate 3$d$ elements the solute-C
interactions are largely affected by the magnetic coupling effects.
For instance, the metal Cr displays a repulsive interaction with C
for all the four configurations. The similar situation has been
observed for Co and Ni. The most striking case is Mn, which is the
only element that shows an unusual character of the binding
interaction with C. The 1$nn$ Mn-C interaction is weakly attractive
(Fig. \ref{fig:fig2}), as accompanied with the appearance of a
ferromagnetic coupling with its 1$nn$ Fe atoms.

The attractive Mn-C interactions were not only studied theoretically
in Refs. \cite{Bakaev,Medvedeva,Numakura}, but also derived from the
experiments \cite{Mn, Mnnew1,Mnnew2,Mnnew3}. Numakura \emph{et
al.}\cite{Numakura} derived the Mn-C interaction energies using the
molecular statics technique based on the empirical pairwise
potentials. Similarly, they also reported the attractive Mn-C
interaction on the 1$nn$ shell but with a bit large binding energy.
This discrepancy might be attributed to their less accurate
empirical pairwise potentials and their bad choice for the energy
reference. In their calculations \cite{Numakura}, they used the
5$nn$ Mn-C interaction configuration as the reference energy,
because they believed that beyond the 5$nn$ shell the Mn and carbon
atoms should not interact each other. However, according to our
calculations, the Mn-C interaction energy at the 5$nn$ configuration
is not negligible (see Fig. \ref{fig:fig6}). Medvedeva \emph{et
al.}\cite{Medvedeva} performed similar first-principles calculations
and they also obtained an attractive Mn-C binding energy (-0.10 eV)
for the 1$nn$ configuration, which is quite accordant with our
result (-0.08 eV). Besides, our results also agree well with the
calculated data recently reported by Bakaev \emph{et
al.}\cite{Bakaev} (see Fig. \ref{fig:fig2}). Although all the
calculated Mn-C binding energies are much lower than the
experimentally estimated values (0.14 eV-0.46 eV) \cite{Mn,
Mnnew1,Mnnew2,Mnnew3}, the theoretical results could better match
the experimental ones if the actual Mn concentration and the
formation of Mn$_x$C clusters were taken into account
\cite{Medvedeva}.

\begin{figure}[b!]
\begin{center}
\includegraphics[width=0.45\textwidth]{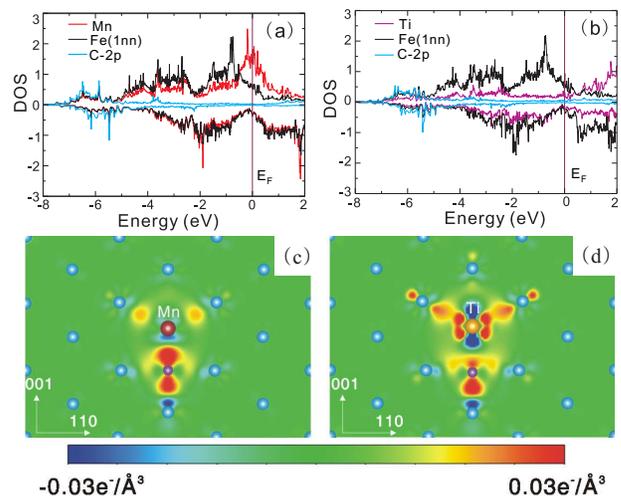}
\end{center}
\caption{Panels (a and b): Local density of states calculated in the
Fe$_{53}M$C ($M$= Mn (a) and Ti  (b))  supercell. Panels (c and d):
Charges density difference map for a single C with (c) Mn and (d) Ti
in the 1$nn$ configuration. The blue balls and purple ball represent
the Fe atoms and the C atom, respectively. The Mn and Ti atoms are
both labeled.}\label{fig:fig4}
\end{figure}

In order to elucidate the attractive Mn-C interaction on the 1$nn$
shell, we analyzed the electronic structures including the charge
density differences and local density of states compared with the
opposite Ti-C binding case. As shown in Fig. \ref{fig:fig4}(a), in
the low energy region from -7 eV to -5 eV of the density of states,
a strong hybridization can be visualized between Mn and C
2\emph{p}-like states, whereas in the Ti-C case the hybridization
between Ti  and C 2\emph{p}-like states is relatively much weaker
(Fig. \ref{fig:fig4}(b)). This fact can be further supported by
their electronic density deformation maps which give a direct
real-space visualization of local electronic rearrangements. It can
be seen from the Fig. \ref{fig:fig4}(c) that the charge accumulation
clearly occurs along the bond between Mn and C in the 1$nn$
configuration, whereas less charges are accumulated between Ti and C
(Fig. \ref{fig:fig4}(d)). In contrast, more charges accumulate
between Ti and its 1$nn$ Fe atoms in the Ti-C case. These results
suggest that Mn atom binds strongly with the C atom rather than the
Fe matrix, whereas Ti atom shows a contrary behavior.

\begin{figure}[b!]
\begin{center}
\includegraphics[width=0.35\textwidth]{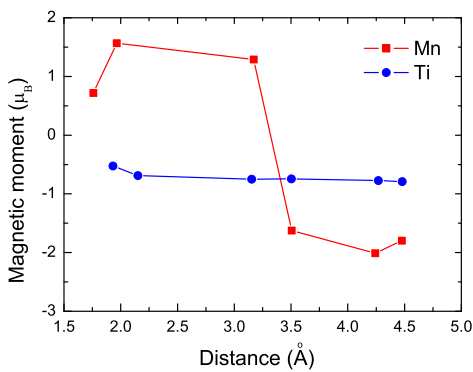}
\end{center}
\caption{The local magnetic moment of the solute $M$ ($M$=Mn and Ti) versus the $M$-C distance in the most
energetically stable Fe$_{53}M$C supercell.
}\label{fig:fig5}
\end{figure}

In agreement with the analysis of electronic structures, it has been
found that the magnetic couplings indeed significantly affect the
Mn-C interactions. On the one hand, the magnetic interactions
decrease the Mn-C binding energy from a positive non-spin-polarized
value of 0.13 eV to a negative spin-polarized value of -0.08 eV,
whereas they almost do not impact the Ti-C binding energy
(spin-polarized, 0.79 eV and non-spin-polarized, 0.77 eV). On the
other hand, we found that the magnetic moment of Mn atom changes
greatly from an antiferromagnetic spin moment of -0.39
$\mu_{\text{B}}$ in Fe$_{53}$Mn supercell to a ferromagnetic spin
moment of 0.72 $\mu_{\text{B}}$ in Fe$_{53}$MnC supercell. This
indicates that the Mn atom is so flexible in the magnetic moment
that it can easily change its sign of the magnetic moment, which has
been confirmed by the Bakaev's work \cite{Bakaev}. However, the
addition of carbon changes a little on the antiferromagnetic coupled
magnetic moment of Ti. Besides, we further derived  the magnetic
moments of the solutes Mn and Ti as a function of the solute-C
distance, as shown in Fig. \ref{fig:fig5}. When the Mn-C distance is
short, the ferromagnetic coupling between Mn and its nearby Fe atoms
is more energetically favorable. However, as the Mn-C distance
increases, Mn prefers the antiferromagnetic coupling with the
neighboring Fe atoms. This fact reveals that carbon can stabilize
the local ferromagnetic coupling between Mn and the neighboring Fe
atoms, in agreement with the Medvedeva's conclusion
\cite{Medvedeva}. However, no obvious change is observed for the
Ti-C case with the increasing Ti-C distance. It thus can be
concluded that the magnetic couplings indeed play an important role
for the abnormal Mn-C interactions.

In fact, there have been numerous efforts to elucidate the influence
of the substitutional atoms on the Snoek peaks measured by the
internal friction experiments
\cite{Saitoh-1,Snoek-1,Snoek-2,Snoek-3}. Interestingly, it has been
observed that the dilute addition (less than 1 mass\%) of solutes
$M$ (Mn, P, Si, Al, Cr, and Co) in bcc Fe-C-$M$ alloys reduces the
normal Snoek peak height, but does not result in the appearance of
any abnormal peaks  \cite{Saitoh-1}. Although the experimental
conditions were far different from our current DFT considerations,
within a qualitative level these experimental facts are still in
agreement with our above analysis of the solute-C binding energies.
Specifically, from our current calculations, almost all of $M$
solutes exhibit large repulsive interactions with carbon in their
1$nn$ and $2nn$ shells. This fact naturally reveals that the normal
regions, where carbon would occupy, should be reduced due to the $M$
addition, leading to the reduction of the normal Snoek peak height.
In contrast, it also needs to be noted that the attractive
interactions between $M$ and C are so weak that they cannot
dramatically increase the number of C atoms in the influenced
regions of the solutes. This fact interprets well the reason as to
why no abnormal peak appears.

\subsection{Solute $M$-vacancy-C interactions}

As mentioned above, the micro-segregation of carbon easily occurs in
the bcc-type steels with the high Mn content \cite{Suzuki,Lu YP}.
Our above calculations for the Mn-C binding interactions seem
consistent with these experimental observations. However, the
attractive interaction between Mn and C at the 1$nn$ configuration
is only -0.08 eV, which is rather weak. Thus, it is hard to believe
that the attractive 1$nn$ Mn-C interaction is the main reason for
the occurrence of carbon micro-segregation in high Mn
steels\cite{Suzuki,Lu YP}.

\begin{figure}[b!]
\begin{center}
\includegraphics[width=0.35\textwidth]{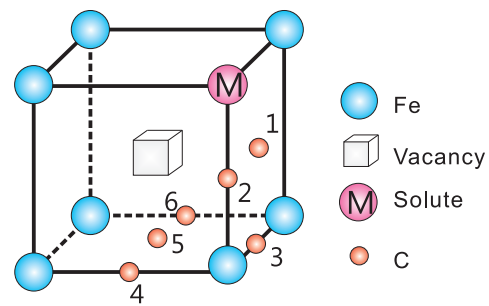}
\end{center}
\caption{Solute $M$-vacancy-carbon configurations. The value $i$
represents the position of the carbon in the $i$th configuration.}
\label{fig:fig7}
\end{figure}

\begin{figure}[b!]
\begin{center}
\includegraphics[width=0.35\textwidth]{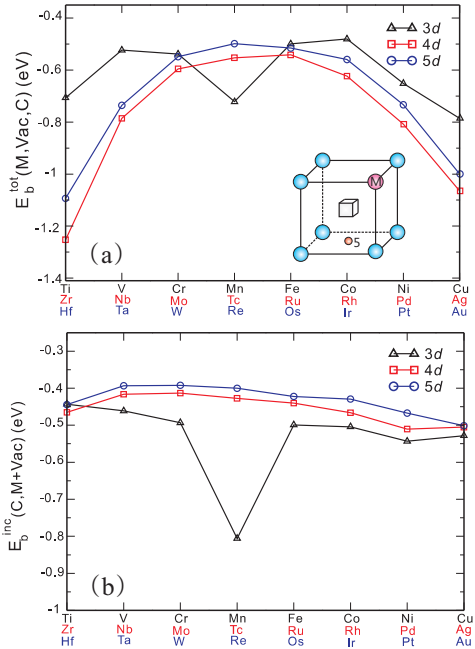}
\end{center}
\caption{(a) The total binding energies among the solute $M$, carbon
and vacancy. (b) The incremental binding energies between the
$M$-vacancy pair and carbon. $M$=3\emph{d} (triangles), 4\emph{d}
(squares), and 5\emph{d} (circles). } \label{fig:fig8}
\end{figure}

Since the vacancy could be easily formed in the bulk, dislocation
core, interfaces and grain boundaries of the steels, we have
attempted to introduce an Fe vacancy to form the so-called
solute-vacancy-C complex. Interestingly, according to several
previous studies \cite{Ohnuma, Olsson, Bakaev} it has been noted
that the Fe vacancy can serve as a strong carbon trapping site due
to the large binding energy between the vacancy and C (-0.59 eV)
\cite{Ohnuma} and the Fe vacancy also shows an attractive ability
to bind the Mn atom with the largest binding energy at the 1$nn$
configuration \cite{Olsson,Bakaev}. Even at the 3$nn$ configuration,
the Mn-vacancy interaction is still attractive \cite{Olsson,Bakaev}.
The systematical calculations further revealed that the vacancy
always exhibits the largest binding energy at the 1$nn$ site with
other $M$ solutes in bcc Fe \cite{Olsson,Bakaev}. Therefore, it can
be inferred that the $M$ solutes and the vacancy can easily form the
solute-vacancy pair in $\alpha$-Fe. Here, based on those most stable
configuration of the solute $M$-vacancy pair, we have further
incorporated a single carbon atom into the $M$-vacancy pair. As
shown in Fig. \ref{fig:fig7}, there are six possible configurations
by taking into account the symmetry. Our calculations demonstrated
that the 5$th$ configuration (called ${cgf}^5$), as illustrated in
the inset of Fig. \ref{fig:fig8}(a), was the most favorable in
energy among the six configurations for all the 3$d$ elements.
Because of the tremendous computing workload, we did not do the test
for the 4$d$ and 5$d$ rows elements. However, one can still
reasonably trust that the ${cgf}^5$ one is also the most stable one
for the 4$d$ and 5$d$ rows elements because in this configuration
both the solute $M$ and carbon are strongest bound to the vacancy
\cite{Ohnuma, Olsson, Bakaev} and the solute $M$-carbon interaction
is less repulsive.

Figure \ref{fig:fig8}(a) shows the total $M$-vacancy-C binding
energies. The negative values indicate that when an Fe vacancy is
introduced, the solute-vacancy-C complex becomes more energetically
stable with respect to the isolated defects. It can thus be inferred
that the solute, vacancy and carbon would easily form a defect
cluster in $\alpha$-Fe. However, when we consider the contribution
of the solute $M$-vacancy pair to the total binding energy
(\emph{i.e}, the incremental binding energy between the solute
$M$-vacancy pair and carbon), the results are much unexpected. As
elucidated in Fig. \ref{fig:fig8}(b), with respect to the single Fe
vacancy all the solute $M$-vacancy pairs for the 4$d$ and 5$d$ rows
elements exhibit weaker binding energies with carbon. Surprisingly,
the Mn-vacancy pair shows a significantly large attractive binding
energy with carbon (about -0.81 eV). This value is nearly twice
larger than those of other $M$-vacancy pairs with carbon (Fig.
\ref{fig:fig8}(b)).

\begin{figure}[b!]
\begin{center}
\includegraphics[width=0.40\textwidth]{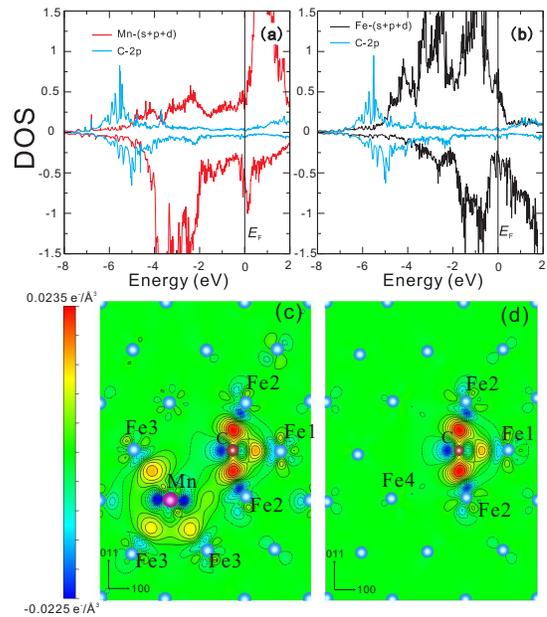}
\end{center}
\caption{Local density of states of the Mn-C interaction (a) and of
the Fe4-C interaction with Fe substituting Mn site (b) in the
\emph{cfg}$^5$ model; their corresponding charge density difference
maps: Mn-vacancy-C (c) and  Fe-vacancy-C (d). The Mn, Fe, and C
atoms are all labeled.} \label{fig:fig9}
\end{figure}

We further compare the local density of states of the Mn-vacancy-C
with those of the Fe-vacancy-C case in bcc Fe in Fig.
\ref{fig:fig9}(a and b). It can be seen that there is no obvious
electronic hybridization between Mn (or Fe4 which just substitutes
the Mn site as marked in Fig. \ref{fig:fig9}(d)) and C. This is
consistent with their charge difference maps where no charge
accumulations are observed between Mn (or Fe4) and carbon. In
contrast, the interstitial carbon atom shows the obvious electronic
hybridizations with its 1$nn$ Fe1 and 2$nn$ Fe2 atoms, as visualized
by the charge accumulations in Fig. \ref{fig:fig9}(c and d). These
comparisons suggest that the anomalous large binding energy between
the Mn-vacancy pair and C does not originate from the electronic
hybridization between Mn and C. Actually, the presence of the Fe
vacancy results in an enhanced spin exchange splitting for Mn. From
Fig. \ref{fig:fig9}(a) the minority spin-down states are mostly
located in the energy range from -4 to -2 eV whereas the majority
spin-up states shift above the Fermi level, thereby causing the
anti-ferromagnetic coupling with its 1$nn$ Fe3 atom with a large
magnetic moment of about 2.67 $\mu_{\text{B}}$  (see Table
\ref{tab:Table1}). This strong magnetic couplings between Mn and
Fe3, as illustrated by the accumulated charges in Fig.
\ref{fig:fig9}(c), play a crucial role in contributing to the
anomalous large binding energy between the Mn-Vacancy pair and C.

\begin{figure}[b!]
\begin{center}
\includegraphics[width=0.45\textwidth]{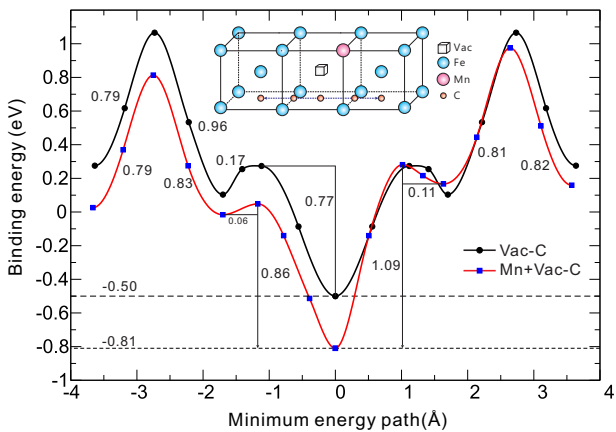}
\end{center}
\caption{Carbon diffusion energy curves in the presence of a single
vacancy or the Mn-vacancy pair in $\alpha$-Fe. The embedded figure
shows the corresponding migration pathways. The values near the
curves represent the energy barrier.} \label{fig:fig10}
\end{figure}

Furthermore, the influences of the Mn-vacancy pair on the carbon
migration, have also been analyzed, as compared with a single Fe
vacancy. In order to eliminate the influence of the mirror images,
the migration energy barriers were calculated using a larger
supercell (4$\times$4$\times$4 bcc unit cell). As shown in Fig.
\ref{fig:fig10}, because of the large vacancy-C attraction on the
1$nn$ shell and repulsion on the 2$nn$ shell \cite{Tapasa K}, the
energy barrier of the carbon atom jumping back towards the vacancy
is far lower than that of the carbon atom escaping away. Hence, we
expect that carbon will return more frequently to the vacancy. A
successful jump is found to occur when the C atom moves directly
from one 1$nn$ $o$-site to another 1$nn$ $o$-site around the vacancy
with the energy barrier of 0.83 eV, nearly the same as that for an
isolated C atom jumping in the defect-free $\alpha$-Fe. This fact
suggests that the carbon motion is restricted within the cell
centered around the vacancy position of the maximum bond with the C
atom. When the Mn-vacancy pair is introduced, carbon will be trapped
more significantly due to their stronger attractive interaction
(-0.81 eV). Compared with a single Fe vacancy, on the one hand, the
energy of Mn-vacancy-C system is much lower, indicating a more
stable state. On the other hand, once the carbon is trapped by the
Mn-vacancy pair, it will be more difficult for carbon atom to escape
from this deeper trap. Even though the carbon jumped to the second
nearest local minima by the thermal fluctuation, it could easily
jump back to the original site because the energy barrier in the
reverse direction is rather small (0.06$\sim$0.11 eV). This fact
reveals that the vacancy assisted by Mn could indeed serve as a
stronger trap to capture the carbon atoms.

Based on the above analysis, it is easily reminiscent of the
occurrence of  carbon micro-segregation in high Mn steels
\cite{Suzuki, Lu YP}. Although the interactions in practice are not
the same as those in our case where the Mn and carbon are dilute, we
can still draw some information from our current calculations for
the first step to understand the nucleation of the carbon
micro-segregation around Mn. Specifically, Mn and vacancy can easily
form a Mn-vacancy pair, not only in the bulk bcc Fe because of their
attractive interactions \cite{Olsson}, but also at the interfaces
due to the attractive interactions \cite{Bakaev} between Mn and the
interfaces where there exists large free volume that might mimic
the presence of the vacancies. Once Mn and vacancy form the pair,
the carbon atoms would be strongly trapped by those Mn-vacancy pairs
due to their large binding energies (see Fig. \ref{fig:fig8}(b)). In
addition, this C-trap is so strong that carbon cannot easily escape
from the vacancy because a large migration energy barrier is
required (see Fig. \ref{fig:fig10}). Both facts make the
Mn-vacancy-C complex stable. Furthermore, it needs to be emphasized
that here we only gave a tentative explanation on the nucleation of
the micro-segregation around Mn. In order to simulate the practical
process, accurate and reasonable Fe-Mn-C potentials should be
developed for larger scale simulations.

\subsection{Influences of the dilute solutes on the carbon's
distribution and chemical potential}

\begin{table}[h]
\footnotesize \caption {Relative positions of carbon at the
different neighboring shells ($nn$) in the 3$\times$3$\times$3 bcc
unit cell. The Solute \emph{M} substitutes Fe at the [000] site in
all the configurations. The positions here are given in units of the
$\alpha$-Fe lattice constant ($\alpha_\text{bcc}$).}
\begin{ruledtabular}
\begin{tabular}{ccccccc}
Shell ($nn$) &1&2&3&4&$5a$&$5b$ \\
C Position &[$\frac{1}{2}$00] &[$\frac{1}{2}$$\frac{1}{2}$0]
&[1$\frac{1}{2}$0] &[1$\frac{1}{2}$$\frac{1}{2}$] &[11$\frac{1}{2}$]
&[$\frac{3}{2}$00] \\
\hline
Shell ($nn$) &6&7&8&9&10&12 \\
C Position &[$\frac{3}{2}$$\frac{1}{2}$0] &[$\frac{3}{2}$10]
&[$\frac{3}{2}$1$\frac{1}{2}$] &[$\frac{3}{2}$11]
&[$\frac{3}{2}$$\frac{3}{2}$0]
&[$\frac{3}{2}$$\frac{3}{2}$1] \\
\end{tabular}
\end{ruledtabular}
\label{tab:Table2}
\end{table}

\begin{figure}[b!]
\begin{center}
\includegraphics[width=0.45\textwidth]{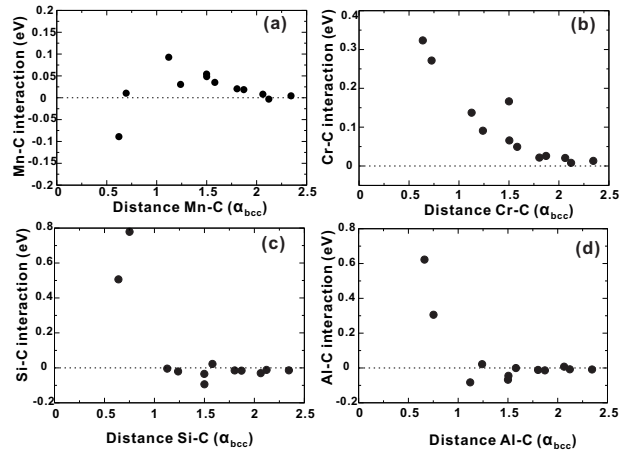}
\end{center}
\caption{Solute $M$-C binding energy versus the $M$-C distance in
units of the bcc lattice constant. $M$=Mn (a), Cr (b), Si (c) and Al
(d). }\label{fig:fig6}
\end{figure}

It is well accepted that substitutional solutes and carbon will
redistribute in the solid solution during the heat treatment and
thermal aging process. To investigate the effects of the solutes on
the carbon's distribution, we further extend the solute-C
interaction range to a farther one since the strain-induced $M$-C
interactions are long-range \cite{Blanter-2, diffusion-2}. As
illustrated in Table \ref{tab:Table2}, the solute $M$ substitutes Fe
atom at the [000] position and the carbon atom varies at twelve
positions up to the 12$nn$ shell with respect to the solute $M$. It
needs to mention that there exist two inequivalent 5$nn$ sites and
no 11$nn$ sites in our current supercell. Here, we only take into
account the solutes Si, Mn, Cr and Al, because these elements are
highly common in ferrite steels and numerous experimental studies
\cite{diffusion-4, Mn,Golovin IS-2,Golovin IS-3,diffusion-1} are
available. As shown in Fig. \ref{fig:fig6}, Cr displays repulsive
interactions with C at all interacting distances, which is also the
case for Mn, except for the 1$nn$ configuration where Mn presents a
weak attractive interaction. However, Si and Al show the different
behaviors. They display the repulsive interactions with C within a
range of one lattice constant but weak attractive interactions
beyond this distance. These computed solute-C interactions are quite
consistent with the experimental observations concerning the
influence of the alloying elements on the carbon solubility limit:
both Mn and Cr hardly change the carbon solubility limit whereas Si
increases it. \cite{Saitoh-2}

\begin{figure}[b!]
\begin{center}
\includegraphics[width=0.45\textwidth]{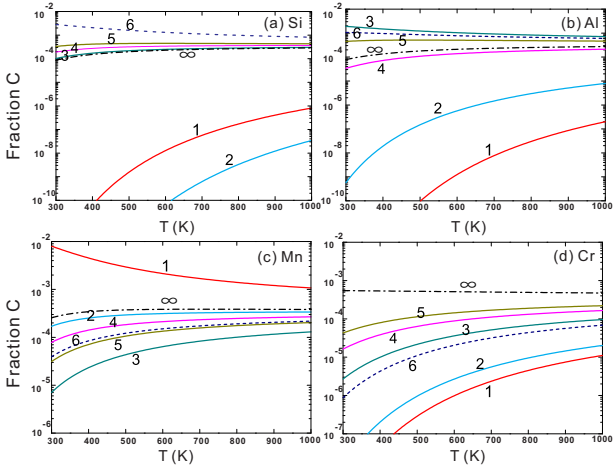}
\end{center}
\caption{Carbon fraction at octahedral interstices for the first
6$nn$ shells with respect to the solute atom \emph{M} (\emph{M} = Si
(a), Al (b), Mn (c), and Cr (d)) and beyond (labeled with $\infty$)
for a composition of Fe$_{0.99}$\emph{M}$_{0.01}$C$_{0.001}$ (Here,
C$_{0.001}$ denotes the fraction of carbon atoms occupying
interstitial octahedral sites).} \label{fig:fig11}
\end{figure}

\begin{figure}[b!]
\begin{center}
\includegraphics[width=0.33\textwidth]{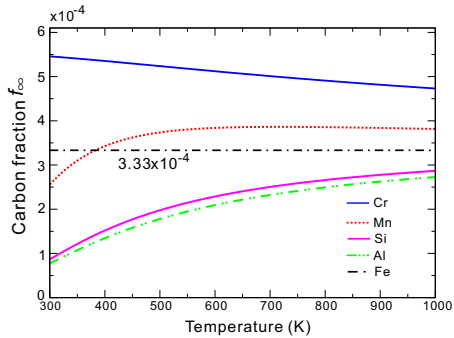}
\end{center}
\caption{Carbon fraction $\emph{f}_{\infty}$ in the octahedral
interstices beyond the solute $M$-C interaction distance as a
function of temperature for a composition of
Fe$_{0.99}M_{0.01}$C$_{0.001}$ ($M$=Mn, Cr, Si and Al).}
\label{fig:fig12}
\end{figure}

In general, within a short distance both the chemical interaction
and strain-induced interaction contribute to the $M$-C binding
energy. The chemical (or attractive) interaction is mostly due to
the electronic structure effect. However, the chemical interaction's
contribution is small here since only a few solutes show weak
attractions with C. In contrast, the strain-induced interactions
between the solutes and C are long-range \cite{Blanter-2,
diffusion-2}. It is strong within short distances and decreases with
the increasing distance. Therefore, it can be inferred that Mn has
relatively large chemical interactions with C since it shows
attractions with C in the 1$nn$ site where the strain should be
large. As the distance rises to a greater one, the chemical
interactions become much weaker and hardly contribute to the $M$-C
binding energy. Then the strain-induced interactions dominate. Thus,
one can expect that the weak attractions between Si (or Al) and C
beyond the distance of one lattice constant are mainly caused by the
strain relief. As the distance further increases beyond the cut-off
radius, the $M$-C binding energy approaches zero.

Based on the thermodynamic considerations, Simonovic \emph{et al}
\cite{Simonovic} proposed a model to analyze how Si affects the
interstitial carbon's distribution and chemical potential in
$\alpha$-Fe. Utilizing the obtained solute-C binding energies, we
further extend the application of this model to other three elements
(Cr, Mn and Al). Figure \ref{fig:fig11} shows the carbon probability
distribution in various neighbor shells around the solute atoms, and
our results reproduce well the results of the carbon interacting
with Si \cite{Simonovic} (Fig. \ref{fig:fig11}(a)). One can see that
there exist significant differences in the carbon distribution for
different solutes. For instance, the carbon atoms can hardly be
found in the 1$nn$ and 2$nn$ shells around Si atom due to their low
fractions which are caused by the large repulsive interactions.
Nevertheless, they would prefer to occupy the interstices between
the 3$nn$ and 6$nn$ shells due to the weak attractions there. Al
acts in a similar fashion. Strikingly, Mn is a unique element that
displays the high carbon fraction in the 1$nn$ site. As for Cr,
within all the shells considered the carbon fractions are lower than
that far beyond Cr.

Figure \ref{fig:fig12} shows the carbon fraction ${f}_{\infty}$ in
the octahedral interstitial sites beyond the solute-C interaction
distance as a function of temperature for a given composition of
Fe$_{0.99}M_{0.01}$C$_{0.001}$. According to the Eq. (\ref{eq:ee7}),
without the solute addition the ${f}_{\infty}$ of the carbon in the
Fe matrix for this composition should be a constant, giving
${f}_{\infty}={C}_{\text{C}}/3=3.33\times10^{-4}$. It can be seen
that the carbon fraction ${f}_{\infty}$ for Mn, Si and Al in the
solid solution increase with the increasing temperature, which is
different for Cr. At all temperatures considered, the carbon
fraction ${f}_{\infty}$ for the Si and Al addition are lower than
that of Fe, whereas it is larger for the Cr addition. Only when the
temperature is above 400 K, will the carbon fraction ${f}_{\infty}$
for the Mn addition exceed that in the pure Fe. Based on the theory
of ideal solutions and in the dilute concentration limit, the
chemical potential of the carbon can be approximated as Eq.
(\ref{eq:ee8}). Accordingly, the carbon¡¯s chemical potential is
positively correlated with the carbon fraction far away from the
solute. From the Fig. \ref{fig:fig12} one can see that Si and Al
additions would decrease the ${f}_{\infty}$, whereas Cr and Mn
increase it, compared with the original one in the matrix without
any solute addition. As a result, it can be inferred that Si and Al
have the potential ability to reduce the carbon's chemical
potential, whereas Mn and Cr increase it.

\subsection{Influences of the solutes on the carbon diffusion}

\begin{table*}
\footnotesize \caption{Diffusion migration energy barriers
$\Delta{E}$ (eV) for the C atom jumping between two nearest neighbor
$o$-sites relative to the solute $M$ ($M$=Mn, Cr, Al and Si). The
positions for the solute $M$ and carbon are shown in the Table
\ref{tab:Table2}.}
\begin{ruledtabular}
\begin{tabular}{ccccccccccc}
&Mn &Cr & Al & Si & &  &Mn &Cr & Al & Si \\
Initial$\rightarrow$Final   &   $\Delta{E}$ &   $\Delta{E}$ &
$\Delta{E}$ &   $\Delta{E}$ &
&   Initial$\rightarrow$Final   &   $\Delta{E}$ &   $\Delta{E}$ &   $\Delta{E}$ &   $\Delta{E}$ \\
\hline
1$\rightarrow$2 &   0.89    &   0.90    &   0.90    &   1.04    &   &   2$\rightarrow$1 &   0.79    &   0.96    &   1.22    &   0.77    \\
2$\rightarrow$3 &   0.94    &   0.77    &   0.64    &   0.59    &   &   3$\rightarrow$2 &   0.86    &   0.90    &   1.03    &   1.38    \\
3$\rightarrow$4 &   0.83    &   0.89    &   1.01    &   0.99    &   &   4$\rightarrow$3 &   0.89    &   0.93    &   0.91    &   1.02    \\
4$\rightarrow$5 &   0.86    &   0.90    &   0.93    &   0.95    &   &   5$\rightarrow$4 &   0.84    &   0.93    &   1.00    &   0.95    \\
5$\rightarrow$6 &   0.92    &   0.81    &   0.96    &   1.03    &   &   6$\rightarrow$5 &   0.95    &   1.04    &   0.82    &   0.80    \\
6$\rightarrow$7 &   0.92    &   0.90    &   0.95    &   0.93    &   &   7$\rightarrow$6 &   0.95    &   0.95    &   0.97    &   1.00    \\
7$\rightarrow$8 &   0.93    &   0.93    &   0.94    &   0.95    &   &   8$\rightarrow$7 &   0.94    &   0.94    &   0.95    &   0.94    \\
8$\rightarrow$9 &   0.91    &   0.91    &   0.95    &   0.92    &   &   9$\rightarrow$8 &   0.93    &   0.93    &   0.91    &   0.95    \\
3$\rightarrow$6 &   0.92    &   0.88    &   0.98    &   0.97    &   &   6$\rightarrow$3 &   0.94    &   0.92    &   0.90    &   0.92    \\
5$\rightarrow$8 &   0.93    &   0.94    &   0.93    &   0.95    &   &   8$\rightarrow$5 &   0.94    &   0.93    &   0.95    &   0.95    \\
7$\rightarrow$10    &   0.93    &   0.92    &   0.95    &   0.94    &   &   10$\rightarrow$7    &   0.98    &   0.93    &   0.96    &   0.96    \\
9$\rightarrow$12    &   0.95    &   0.94    &   0.93    &   0.95    &   &   12$\rightarrow$9    &   0.95    &   0.93    &   0.98    &   0.95    \\
\end{tabular}
\end{ruledtabular}
\label{tab:Table3}
\end{table*}

\begin{figure}[b!]
\begin{center}
\includegraphics[width=0.35\textwidth]{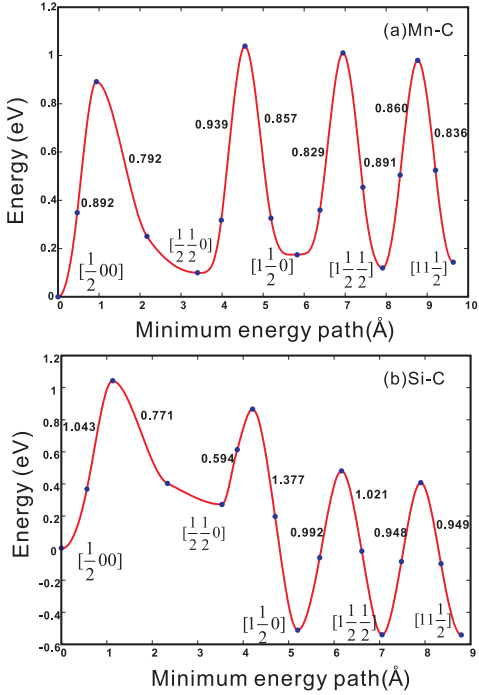}
\end{center}
\caption{The minimum-energy paths of the carbon migrating from the
1$nn$ to the 5$nn$ $o$-sites with respect to the solute Mn (a) or Si
(b) in $\alpha$-Fe. The values with and without square brackets
 represent the carbon's positions
 (see Table \ref{tab:Table2}) and the energy barriers, respectively.}
\label{fig:fig13}
\end{figure}

The most likely and intuitive jumping mechanism for the interstitial
diffusion of the carbon in the $\alpha$-Fe lattice is the jump from
an octahedral site (\emph{o}-site) to another nearest neighboring
one via the tetragonal site (\emph{t}-site) \cite{Jiang-1}. We have
calculated the carbon diffusion migration energy barrier of 0.89 eV
in the pure $\alpha$-Fe, in good agreement with other DFT results of
0.86 eV\cite{Jiang-1} and 0.92 eV \cite{Domain,Domain C}, and the
experimentally measured data of 0.87 eV\cite{Wert}, 0.88 eV
\cite{Takaki} and 0.84 eV \cite{carbides-1}. Moreover, the diffusion
prefactor $D_0$ have been further derived by DFT calculations
according to the formula of $D_0=\frac{1}{6} \alpha
^2(\prod_{j=1}^{3N}\upsilon_j^{ini}/\prod_{j=1}^{3N-1}\upsilon_j^{sad})$
\cite{Jiang-1}, where $\alpha$ is the lattice constant, and
$\upsilon_j^{ini}$ and $\upsilon_j^{sad}$ are the normal-mode
frequencies at the initial and saddle-point state, respectively. Our
result yields $D_0 = 1.56 \times 10^{-7} m^2/s$. This value also
agrees well with the previously calculated data of $1.44 \times
10^{-7} m^2/s$ $\ $\cite{Jiang-1} and $1.66 \times 10^{-7} m^2/s$ $\
$\cite{Simonovic}, and the experimentally determined value of $1.67
\times 10^{-7} m^2/s$ below 350 K \cite{Silva}. In addition, we also
used the kMC simulations to estimate $D_0$ for the C diffusion,
obtaining a value of $2.14 \times 10^{-7} m^2/s$ by fitting the
computed C diffusivities to the temperatures according to Eq.
(\ref{eq:ee11}), in nice agreement with above results.

\begin{figure}
\begin{center}
\includegraphics[width=0.4\textwidth]{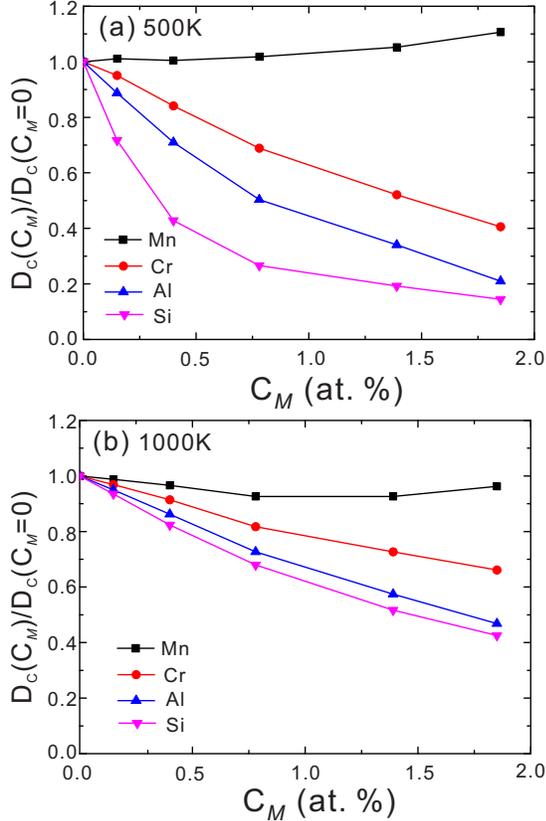}
\end{center}
\caption{Carbon diffusivities in the presence of the solute $M$
($M$=Mn, Cr, Al and Si) relative to that without $M$ versus the
solute concentrations at temperatures of 500 and 1000 K. }
\label{fig:fig14}
\end{figure}

Next we focus on the effects of the dilute substitutional solutes
$M$ ($M$=Mn, Cr, Al and Si) on the carbon's diffusion in
$\alpha$-Fe. We have considered all the possible diffusion pathways
for the carbon jumping between two nearest neighboring $o$-sites in
the presence of solute $M$ using the CI-NEB method \cite{NEB,NEB-2}.
The calculated diffusion migration energy barriers $\Delta{E}$ were
compiled in Table \ref{tab:Table3}. As an example, we further
plotted the minimum-energy paths of the carbon migrating from the
1$nn$ site to the 5$nn$ site with respect to the solute Mn or Si in
Fig. \ref{fig:fig13}. Coupling these DFT energy barriers with the
kMC simulations, the carbon diffusivity and effective migration
energy barrier affected by the solute $M$ have been computed at
various solute concentrations and temperatures, as compiled in Table
\ref{tab:Table4}. The solute concentration dependent trends were
further presented in Fig. \ref{fig:fig14}.

\begin{table*}
\footnotesize \caption{kMC simulated carbon diffusivity $D$ (in
m$^2$/s) and effective migration energy $\Delta{E}$  (in eV) in
$\alpha$-Fe as a function of the solute concentration $C_M$ (in
at.\%) ($M$=Mn, Cr, Al and Si) at temperatures of 500 K and 1000 K.
}
\begin{ruledtabular}
\begin{tabular}{ccccccccccccc}
 &\multicolumn{3}{c}{Mn}   & \multicolumn{3}{c}{Cr} & \multicolumn{3}{c}{Al} & \multicolumn{3}{c}{Si}  \\
\cline{2-4} \cline{5-7} \cline{8-10} \cline{11-13}
$C_\text{\emph{M}}$    &   500 K   &   1000 K  &   $\Delta{E}$ &   500 K   &   1000 K  &   $\Delta{E}$ &   500 K   &   1000 K  &   $\Delta{E}$ &   500 K   &   1000 K  &   $\Delta{E}$ \\
(at.\%) & ($10^{-16}$) & ($10^{-12}$) & (eV)      & ($10^{-16}$) & ($10^{-12}$) & (eV)   & ($10^{-16}$) & ($10^{-12}$) & (eV)    & ($10^{-16}$) & ($10^{-12}$) & (eV)  \\
\hline
0.00    &   2.55    &   7.38    &   0.89    &   2.55    &   7.38    &   0.89    &   2.55    &   7.38    &   0.89    &   2.55    &   7.38    &   0.89    \\
0.15    &   2.59    &   7.30    &   0.88    &   2.43    &   7.15    &   0.89    &   2.27    &   7.00    &   0.89    &   1.83    &   6.90    &   0.91    \\
0.40    &   2.58    &   7.13    &   0.88    &   2.15    &   6.75    &   0.89    &   1.81    &   6.35    &   0.90    &   1.09    &   6.08    &   0.94    \\
0.78    &   2.61    &   6.83    &   0.88    &   1.76    &   6.03    &   0.90    &   1.29    &   5.35    &   0.92    &   0.68    &   5.03    &   0.94    \\
1.39    &   2.68    &   6.83    &   0.86    &   1.33    &   5.35    &   0.91    &   0.87    &   4.25    &   0.93    &   0.49    &   3.83    &   0.97    \\
1.85    &   2.83    &   7.10    &   0.87    &   1.04    &   4.88    &   0.93    &   0.54    &   3.45    &   0.95    &   0.37    &   3.15    &   0.98    \\
\end{tabular}
\end{ruledtabular}
\label{tab:Table4}
\end{table*}

At first, it needs to be emphasized that we have reproduced well the
results of the C diffusivity in the presence of Si obtained by
Simonovic \emph{et al.}\cite{Simonovic}. Our data show the same
order of magnitude and the similar tendency as theirs, but ours are
much smaller, in particular at the low temperature. The discrepancy
is likely due to the different migration energies used for the C
diffusion. In our calculations, we considered the
direction-dependent diffusion energy barriers. However, in Ref.
\cite{Simonovic} the kinetically resolved migration barrier
(KRA-approximation) proposed by Van der Ven \emph{et al.}\cite {Van
der Ven} has been employed. It was defined as the average of the
forward and backward diffusion migration energies to overcome the
difficulties associated with the direction dependence of the
diffusion migration energy. It is really a good approximation only
when the thermodynamic energy difference before and after the C jump
is much smaller than the corresponding kinetic parts \cite{Van der
Ven}. This kinetic part was assumed to be equal to the migration
energy for the C diffusion at the infinite distance away from the
solute atom. However, from our calculations it was found that for
the Si addition case the thermodynamic part has the same order of
magnitude with the kinetic part within the 3$nn$ shells, disobeying
this KRA-approximation \cite{Van der Ven}. Therefore, it would be
more reasonable to fully take into account the direction-dependent
diffusion energy barriers for each jump in computing the carbon
diffusivity.

Furthermore, it can be found from Fig. \ref{fig:fig14} that, in the
range of dilute solute concentration Mn exhibits little influence on
the carbon diffusivity (because it does not significantly alter the
carbon diffusion energy barriers), whereas Cr, Al and Si all
remarkably reduce the carbon diffusivity, particularly at low
temperature (\emph{i.e.}, 500 K). Taking Si for instance, the 0.78
at.\% content of Si significantly reduces the C diffusivity with
73\% at 500 K and 32\% at 1000 K than those without the Si addition.
This is due to the fact that Si greatly affects the migration energy
barriers of the carbon diffusion, as shown in Fig. \ref{fig:fig13}.
Specifically, the 1$nn$ and 2$nn$ $o$-sites around Si are so high in
energy that it is more difficult for carbon to stay, thereby
significantly reducing the positions where the carbon can diffuse.
At the longer distance beyond the 3$nn$ shells, the carbon atom is
trapped by Si with a weak attraction, which increases the residence
time of carbon to stay at these sites. Both of the above situations
would contribute to the significant reduction of the carbon
diffusion. The similar behavior is also observed for the Al
addition. As for Cr, its reduction on the C diffusivity is less
apparent since only labyrinth mechanism \cite{Simonovic} works due
to its repulsive interactions with C in all the twelve nearest
neighboring shells.

Our simulated results also demonstrate that, with increasing the
solute concentration of Cr (Al or Si), the C diffusivity decreases
significantly (see Fig. \ref{fig:fig14}). Interestingly, at low
temperature the Si (or Al) addition makes carbon atoms most likely
sit at the attractive-interaction region around the solute,
remarkably decreasing carbon diffusivity (\emph{i.e.}, 500 K in Fig.
\ref{fig:fig14}). It indicates that the carbon diffusion is indeed
dominated by the so-called trapping mechanism \cite{traping} at the
low temperature. However, the high-temperature kMC simulations
reveal that the carbon atom would randomly occupy any interstitial
\emph{o}-sites. In this situation, both the labyrinth mechanism
\cite{Simonovic} and trapping mechanism \cite{traping} work well. As
evidenced in our simulations, at high temperature (\emph{i.e.}, 1000
K in Fig. \ref{fig:fig14}) the alloying addition results in a less
impact on the carbon diffusivity.

The influences of the Al concentration on the carbon diffusion in
Fe-Al-C alloys have also been investigated experimentally by Strahl
and Golovin \emph{et al.} \cite{diffusion-1}. They observed that
with increasing the Al content the Snoek peak became broader and its
corresponding position shifted upward the higher temperature.
Similar phenomenon was also observed in the Fe-Cr-C alloys
\cite{Blanter-1}. It is known that the Snoek peak is a typical
relaxational internal friction peak due to the migration of the
interstitial atoms induced by the stress. In the bcc Fe-C alloys
without any solutes, where all the $o$-sites are equivalent for the
carbon, the individual migration energy barrier and effective
migration energy barrier $\Delta{E}$ are the same for each jump of
carbon. Thus, the relaxation time $\tau$ ($\tau=\tau_0
exp(\Delta{E}/k_BT)$) should be nearly constant, i.e., the resonance
condition is almost strictly satisfied, leading to the appearance of
the narrow Snoek peak. However, when Al is added, it would greatly
affect the potential energy surface and the distribution of the
interstitial carbon atoms, which makes the resonance condition in
the internal friction measurements less strictly to meet. That is
the reason why the Snoek peak broadens when Al is added. The center
of the wide peak corresponds to the effective migration energy
barrier of the carbon. Within our kMC simulations it was also found
that with increasing the Al content, the effective migration energy
barrier of the carbon diffusion increases (see Table
\ref{tab:Table4}). This fact indicates that a much higher
temperature is required to activate the migration of the carbon,
coinciding well with the experimentally observed upper shift of the
Snoek peak to the higher temperature \cite{diffusion-1}.

\section{Conclusions}

In the present paper, we have systematically investigated the dilute
solute-C, solute-vacancy-C interactions and the influences of dilute
solutes on the carbon's distribution and diffusion in $\alpha$-Fe
through first-principles calculations. The main conclusions are as
follows:

($i$) In terms of the 4$d$ and 5$d$ elements, the solute-C
interactions are mostly governed by the strain relief, whereas for
the 3$d$ elements, magnetic coupling and electronic structure also
play important roles, which may override the strain relief. Mn is
the only element that shows attractive interactions with C in the
1$nn$ shell.

($ii$) When an Fe vacancy is introduced, the solute-vacancy-carbon
total binding energies become negative, indicating the easy
formation of the defects complex. In particular, the Mn-vacancy pair
exhibits an exceptionally large binding energy with C (-0.81 eV),
which is due to the stronger anti-ferromagnetic coupling between Mn
and its nearest neighboring Fe atoms assisted by the Fe vacancy.
Moreover, our results also suggest that the vacancy assisted by Mn
could serve as a stronger trap site to the carbon.

($iii$) The longer-range interactions between the dilute solutes
(Mn, Cr, Al and Si) and C have been investigated in details. Through
the model proposed by Simonovic\cite{Simonovic} coupled with the
thermodynamic considerations, it has been found that the solutes
addition would greatly affect the carbon's distribution and chemical
potential. Among the four solutes, Mn and Cr tend to increase the
carbon's chemical potential whereas Al and Si reduce it.

($iv$) The carbon diffusion affected by the dilute solutes has been
modeled in-depth through the kinetic Monte Carlo simulations coupled
with the DFT energy barrier calculations. The results demonstrate
that in the range of the dilute concentration, Mn hardly changes the
carbon diffusivity whereas Cr, Al and Si significantly decrease it
as the concentration increases.

Finally, we would like to emphasize that our current
first-principles calculations only fit to the cases where the
alloying solutes and carbon are dilute in $\alpha$-Fe. When compared
with the available experimental results, one should be always
cautious about complicated experimental factors, such as
temperature, pressure, concentration, and defects, and so on.

\section*{Acknowledgements}

This work was supported by the "Hundred Talents Project" of the
Chinese Academy of Sciences and from the Major Research Plan (Grand
Number: 91226204) of the NSFC of China (Grand Numbers:51174188 and
51074151) and Beijing Supercomputing Center of CAS (including its
Shenyang branch).


\begin{thebibliography}{99}

\bibitem{Abbaschian R}  R. Abbaschian, L. Abbaschian, and R. E. Reed-Hill, \emph{Physical Metallurgy Principles} (Stamford, Cengage Learning, 2009).
\bibitem{Saitoh-1} H. Saitoh, N. Yoshinaga, and K. Ushioda, Acta Mater. \textbf{52}, 1255 (2004).
\bibitem{Saitoh-2} H. Saitoh, K. Ushioda, N. Yoshinaga, and W. Yamada, Scripta Mater. \textbf{65}, 887 (2011).
\bibitem{Suzuki} K. Suzuki and T. Miyamoto, Trans.  ISIJ \textbf{18}, 80 (1978).
\bibitem{Lu YP} Y. P. Lu, R. F. Zhu, S. T. Li, F. C. Zhang, and S. Q. Wang, Prog. Nature Sci. \textbf{40}, 567 (1996).
\bibitem{diffusion-1} A. Strahl, I. S. Golovin, H. Neuh\"{a}user, S. B. Golovina, and H. R. Sinning, Mater. Sci. Eng. A \textbf{442}, 128 (2006).
\bibitem{diffusion-2} M. S. Blanter and L. B. Magalas, Scripta Mater. \textbf{43}, 435 (2000).
\bibitem{diffusion-3}  I. S. Golovin, S. V. Divinski, J. \v{C}\'{\i}\v{z}ek, I. Proch\'{a}zka, and F. Stein, Acta Mater. \textbf{53}, 2581 (2005).
\bibitem{diffusion-4} D. Ruiz, J. L. Rivera-Tovar, D. Segers, R. E. Vandenberghe, and Y. Houbaert, Mater. Sci. Eng. A \textbf{442}, 462 (2006).
\bibitem{carbides-1}  F. Walz, T. Wakisaka, and H. Kronm\"{u}ller, Phys. Status Solidi A \textbf{202}, 2667 (2005).
\bibitem{carbides-2}  S. Garruchet and M. Perez, Comput. Mater. Sci. \textbf{43}, 286 (2008).
\bibitem{carbides-3}  C. K. Ande and M. H. F. Sluiter, Metall. Mater. Trans. A \textbf{43}, 4436 (2012).
\bibitem{Sozinov}  A. G. Khachaturyan, \emph{Theory of Structural Transformations in Solids.} (Wiley, New York, 1983).
\bibitem{Jiang-1} D. Jiang and E. Carter, Phys. Rev. B  \textbf{67}, 214103 (2003).
\bibitem{Jiang-2} D. Jiang and E. Carter, Phys. Rev. B  \textbf{71}, 045402 (2005).
\bibitem{Domain} C. Domain and J. Foct, Phys. Rev. B \textbf{69}, 144112 (2004).
\bibitem{Ohnuma} T. Ohnuma, N. Soneda, and M. Iwasawa, Acta Mater. \textbf{57}, 5947 (2009).
\bibitem{Yan} Y. You, M. F. Yan, and H. T. Chen, Comput. Mater. Sci. \textbf{67}, 222 (2013).
\bibitem{Becquart} C. S. Becquart, J. M. Raulot, G. Bencteux, C. Domain, M. Perez, S. Garruchet, and H. Nguyen, Comput. Mater. Sci. \textbf{40}, 119 (2007).
\bibitem{Simonovic} D. Simonovic and C. K. Ande, Phys. Rev. B \textbf{81}, 054116 (2010).
\bibitem{Bakaev} A. Bakaev, D. Terentyev, G. Bonny, T. P. C. Klaver, P. Olsson, and D. Van Neck, J. Nucl. Mater.  \textbf{444}, 237 (2014).
\bibitem{Hohenberg} P. Hohenberg, Phys. Rev. B \textbf{136}, 864 (1964).
\bibitem{Kohn} W. Kohn and L. J. Sham, Phys. Rev. A \textbf{140}, 1133 (1965).
\bibitem{Kresse-1}  G. Kresse and J. Hafner, Phys. Rev. B \textbf{47}, 558 (1993).
\bibitem{Kresse-2}  G. Kresse and J. Furthm\"{u}ller, Phys. Rev. B \textbf{54}, (11169) 1996.
\bibitem{PAW}  P. E. Bl\"{o}chl, Phys. Rev. B \textbf{50}, 17953 (1994).
\bibitem{PBE}  J. P. Perdew, K. Burke, and M. Ernzerhof, Phys. Rev. Lett. \textbf{77}, 3865 (1996).
\bibitem{Singh} D. J. Singh, W. E. Pickett, and H. Krakauer, Phys. Rev. B \textbf{43}, 11628 (1991).
\bibitem{Klymko} T. Klymko and M. H. F. Sluiter, J. Mater. Sci. \textbf{47}, 7601 (2012).
\bibitem{Counts} W. A. Counts, C. Wolverton, and R. Gibala, Acta Mater. \textbf{58}, 4730 (2010).
\bibitem{Kittel C}  C. Kittel, \emph{Introduction to solid state physics} (Wiley, New York, 1996).
\bibitem{NEB} G. Henkelman, B. P. Uberuaga, and H. Jonsson, J. Chem. Phys. \textbf{113}, 9901 (2000).
\bibitem{NEB-2} D. Sheppard, R. Terrell, and G. Henkelman, J. Chem. Phys. \textbf{128}, 134106 (2008).
\bibitem{Kalos} M. H. Kalos and P. A. Whitlock, \emph{Monte Carlo Methods Volume 1: Basics} (Wiley, New York, 1986).
\bibitem{Pascheto} W. Pascheto and G. P. Johari, Metall. Mater. Trans. A \textbf{27}, 2461 (1996).
\bibitem{Mehrer} M. L\"{u}bbehusen and H. Mehrer, Acta Metall Mater \textbf{38}, 283(1990).
\bibitem{Huang} S. Huang, D. L. Worthington, M. Asta, V. Ozolins, G. Ghosh, and P. K. Liaw, Acta Mater. \textbf{58}, 1982 (2010).
\bibitem{TST} J. C. Christopher, \emph{Essential of Computational Chemistry:  Theories and Models} (Wiley, England, 2002).
\bibitem{residence-time algorithm} A. B. Bortz, M. H. Kalos, and J. L. Lebowitz, J. Comput. Phys. \textbf{17}, 10 (1975).
\bibitem{Einstein} H. Mehrer, \emph{Diffusion in Solids: Fundamentals, Methods, Materials, Diffusion-Controlled Processes} (Springer, Germany, 2007).
\bibitem{Olsson} P. Olsson, T. P. C. Klaver, and C. Domain, Phys. Rev. B \textbf{81}, 054102 (2010).
\bibitem{Sandberg} N. Sandberg, K. Henriksson, and J. Wallenius, Phys. Rev. B \textbf{78}, 094110 (2008).
\bibitem{Medvedeva} N. I. Medvedeva, D. C. Van Aken, and J. E. Medvedeva, J. Phys.: Condens. Matter \textbf{23}, 326003 (2011).
\bibitem{Numakura} H. Numakura, G. Yotsuit, and M. Koiwa, Acta metall mater. \textbf{43}, 705 (1995).
\bibitem{Mn} V. Massardier, E. L. Patezour, M. Soler, and J. Merlin, Metall. Mater. Trans. A \textbf{36}, 1745 (2005).
\bibitem{Mnnew1} H. Abe, T. Suzuki, and S. Okada, Trans. Jpn. Inst. Met. \textbf{25}, 215 (1984).
\bibitem{Mnnew2} H. Abe, J. Korean Inst. Met. \textbf{24}, 612 (1986).
\bibitem{Mnnew3} T. Nishizawa, Bull. Jpn. Inst. Met. \textbf{12}, 401 (1973).
\bibitem{Snoek-1}  C. Wert, J. Metals \textbf{4}, 602 (1952).
\bibitem{Snoek-2} H. Numakura and M. Koiwa, \emph{International Symposium, M3D, Mechanics and Mechanisms of Materials Damping ASTM} 383 (1997).
\bibitem{Snoek-3} H. Numakura and M. Koiwa, J Phys IV Coll C \textbf{8}, 97 (1996).
\bibitem{Blanter-2} M. S. Blanter, J. Alloys Compd. \textbf{291}, 167 (1999).
\bibitem{Golovin IS-2} I. S. Golovin IS, H. Neuh\"{a}user, A. Rivi\`{e}re, and A. Strahl, Intermetallics \textbf{12}, 125 (2004).
\bibitem{Golovin IS-3} I. S. Golovin and S. B. Golovina, The Physics of Metals and Metallography \textbf{102}, 593 (2006).
\bibitem{Tapasa K} K. Tapasa, A. Barashev, D. Bacon, and Y. Osetsky, Acta Mater. \textbf{55}, 1 (2007).
\bibitem{Domain C} C. Domain, J. Nucl. Mater. \textbf{351}, 1 (2006).
\bibitem{Wert} C. A. Wert, Phys. Rev. \textbf{79}, 601 (1950).
\bibitem{Takaki} S. Takaki, J. Fuss, H. Kugler, U. Dedek, and H. Schults, Radiat. Eff.  \textbf{79}, 87 (1983).
\bibitem{Silva} J.R.G. da Silva and R.B. McLellan, Mater. Sci. Eng. \textbf{26}, 83 (1976).
\bibitem{Van der Ven} A. Van der Ven, G. Ceder, M. Asta, and P. Tepesch, Phys. Rev. B \textbf{64}, 184307 (2001).
\bibitem{traping} C. A. Wert and R. C. Frank, Annu. Rev. Mater. Sci. \textbf{13}, 139 (1983).
\bibitem{Blanter-1}  I. S. Golovin, M. S. Blanter, and R. Schaller, phys. stat. sol. A \textbf{160}, 49 (1997).
\end{thebibliography}
\end{document}